\documentclass[lettersize,journal]{IEEEtran}
\IEEEoverridecommandlockouts

\usepackage{cite}
\usepackage{amsmath,amssymb,amsfonts}
\usepackage[ruled,vlined]{algorithm2e}
\SetKw{KwOut}{Output:}
\usepackage{algorithmic}
\usepackage{graphicx}
\usepackage{textcomp}
\usepackage{xcolor}
\usepackage{svg}
\usepackage{listings}
\usepackage{tabularx}
\usepackage{balance}

\definecolor{codebg}{RGB}{248,248,248}
\definecolor{coderule}{RGB}{210,210,210}
\definecolor{scaleranblue}{RGB}{25,90,170}

\lstdefinestyle{scaleran}{
    language=Python,
    basicstyle=\ttfamily\small,
    backgroundcolor=\color{codebg},
    frame=single,
    rulecolor=\color{coderule},
    framerule=0.5pt,
    showstringspaces=false,
    columns=fullflexible,
    keepspaces=true,
    commentstyle=\itshape,
    escapeinside={(*@}{@*)}
}

\begin{document}

\title{ScaleRAN: A Runtime for Elastic GPU Resource Management in AI-RAN}

\author{
 \IEEEauthorblockA{Arthur Hamard$^{*}$, J\'er\'emie Leguay$^{*}$, Gopalasingham Aravinthan$^{\dagger}$, 
 Alberto Conte$^{*}$, \\Julien Lallet$^{*}$, Abdelkader Outtagarts$^{*}$\\$^{*}$Nokia Bell Labs, $^{\dagger}$Nokia Mobile Networks, France}
 \thanks{This work has been submitted to the IEEE for possible publication.}\vspace{-0.8cm}}

\maketitle

\begin{abstract}
AI-RAN is driving the convergence of radio access network (RAN) processing and artificial intelligence (AI) workloads on GPU-enabled computing platforms, ranging from GPU-accelerated base stations to cloud-native infrastructures. While GPU acceleration improves the efficiency of virtualized RAN functions, the computational demand of latency-critical baseband processing varies with traffic load, creating transient spare GPU capacity that can be reclaimed by co-located AI applications. Existing GPU-sharing mechanisms rely on coarse-grained hardware partitioning or static resource allocation, limiting their ability to adapt to slot-level workload fluctuations while preserving strict timing requirements.

This paper presents ScaleRAN, a node-local runtime for elastic GPU resource management in AI-RAN. ScaleRAN leverages NVIDIA Green Contexts to dynamically partition Streaming Multiprocessors (SMs) between RAN and AI workloads through a ladder of pre-created GPU partitions, enabling low-overhead resource reallocation at RAN slot granularity. The proposed runtime naturally supports  Open-RAN (O-RAN) O-Cloud deployments, one representative AI-RAN deployment scenario. To drive these reallocations, we propose a reactive controller inspired by Kubernetes autoscaling and a proactive controller based on quantile regression and conformal prediction. We also develop LDPCSim, a GPU-accelerated workload generator built on NVIDIA Aerial for reproducible evaluation. Results show that predictive fine-grained GPU partitioning satisfies slot-level RAN deadlines while increasing the GPU resources available to co-located AI workloads, demonstrating the practicality of elastic GPU sharing for future AI-RAN platforms.
\end{abstract}


\section{Introduction}
\label{sec:intro]}


\IEEEPARstart{M}{obile} networks are undergoing a fundamental transformation toward AI-RAN~\cite{kundu2025ai}, where radio access network (RAN) processing and artificial intelligence (AI) workloads increasingly converge on GPU-enabled computing platforms. This evolution spans GPU-accelerated base stations and cloud-native deployments, driven by the adoption of commodity GPUs for virtualized RAN functions. In parallel, the O-RAN Alliance~\cite{polese2023understanding} is promoting an open and disaggregated RAN architecture, where radio functions are virtualized and deployed on cloud-native O-Cloud infrastructures. More recently, O-RAN WG6 (Cloudification and Orchestration) has extended this vision toward heterogeneous O-Cloud platforms hosting both virtualized RAN functions and AI applications, including its ongoing work on \emph{O-Cloud Enhancements for AI Applications}~\cite{oran_wg6}. O-Cloud will therefore represent an important AI-RAN deployment platform, providing a standardized cloud-native environment. In this context, GPUs are emerging as a common acceleration platform for virtualized 5G and future 6G RAN deployments, enabling gNBs and edge nodes to support communication and AI workloads. Beyond accelerating latency-critical RAN processing, these platforms can host AI inference and training workloads for AI-for-RAN applications while enabling AI-and-RAN and AI-on-RAN deployment models. This convergence calls for intelligent resource management mechanisms to dynamically share GPU resources between RAN and AI workloads.

As GPUs represent high-value and scarce accelerator resources, maximizing their utilization is essential to improve the efficiency and economic viability of AI-RAN deployments. Rather than dedicating GPUs exclusively to RAN acceleration, AI-RAN platforms will expose GPUs as shared computing resources. This paradigm is relevant to cloud-native O-Cloud deployments, where virtualized RAN and AI workloads are co-located. When the RAN traffic load decreases, the computational demand of processing-intensive functions such as Low-Density Parity-Check (LDPC)~\cite{qi2026decodex} encoding and decoding is reduced, allowing otherwise idle GPU resources to be reclaimed for AI inference or training tasks. Conversely, during periods of increased network load, RAN processing must be prioritized. The challenge is therefore to exploit transient GPU slack without compromising deterministic RAN performance.

Recent work has investigated NVIDIA GPU-sharing technologies for spatial and temporal partitioning to improve utilization under heterogeneous workloads. Multi-Process Service (MPS)~\cite{MPS} enables concurrent execution across multiple CUDA processes, but provides limited isolation and coarse-grained resource control. In contrast, Multi-Instance GPU (MIG)~\cite{MIG} provides hardware-enforced spatial partitioning by dividing the GPU into isolated instances, and is often combined with MPS to improve utilization. Building upon these primitives, hybrid systems~\cite{zhang2024miger,lee2024parvagpu} jointly optimize MIG partition sizing and MPS-based concurrency to balance isolation and efficiency, while kernel- and runtime-level approaches~\cite{jiang2026mmk,zhao2025tally} leverage kernel scheduling and CUDA API interception to mitigate interference among co-located workloads. More recently, Green Contexts (GCs)~\cite{GC,radig2025draco} have introduced a lightweight driver-level abstraction for fine-grained spatial partitioning at the granularity of Streaming Multiprocessors (SMs), enabling dynamic GPU resource allocation.

Collectively, these mechanisms provide complementary trade-offs between isolation, granularity, and runtime flexibility. However, efficiently combining strict RAN latency guarantees, with transmission slots as short as 0.5 ms for commonly used 5G numerologies, and fine-grained, low-overhead reallocation of GPU resources to co-located AI workloads remains an open challenge. Despite growing interest in AI-RAN co-location, most existing works still rely on coarse-grained GPU partitioning, primarily based on MIG~\cite{yet2026sla,shah2025proactive}, or relatively static partitioning and offline optimization~\cite{zhang2024miger,lee2024parvagpu}. As a result, they cannot efficiently exploit transient GPU slack under fluctuating RAN load. These limitations motivate a unified, fine-grained runtime that dynamically reallocates GPU resources between RAN and AI workloads in real time.

This paper makes three main contributions. First, we introduce ScaleRAN, a node-local runtime for elastic GPU resource management that dynamically reallocates SMs between latency-critical RAN and co-located AI workloads through Green Context-based SM partitioning. Second, we propose both reactive and prediction-based autoscaling policies that dynamically determine the minimum GPU allocation required to satisfy slot-level RAN deadlines. Third, we develop LDPCSim, a realistic GPU-accelerated 5G NR workload generator built on NVIDIA Aerial that enables reproducible evaluation of elastic GPU resource management for AI-RAN. Using this platform, we implement and evaluate ScaleRAN, showing that predictive fine-grained GPU partitioning preserves RAN timing guarantees while substantially improving GPU utilization for co-located AI workloads.

The remainder of this paper is organized as follows. Sec.~\ref{sec:related} reviews GPU-sharing mechanisms and AI-RAN resource management. Sec.~\ref{sec:workloads} characterizes the RAN and AI workloads driving elastic GPU allocation. Sec.~\ref{sec:autoscaler} presents the design, implementation, and O-RAN integration of ScaleRAN. Sec.~\ref{sec:policies} introduces the reactive and proactive autoscaling policies, and Sec.~\ref{sec:eval} evaluates them under realistic AI-RAN workloads. Finally, Sec.~\ref{sec:conclusion} concludes the paper.

\section{Related Work}
\label{sec:related}

This section reviews GPU resource-sharing mechanisms most relevant to this work. Since ScaleRAN targets NVIDIA GPU platforms, the discussion focuses on their  technologies.

\subsection{Background on GPU Execution Model } 
\label{sec:gpu-model}

Modern NVIDIA GPUs follow a heterogeneous execution model in which the CPU launches compute functions, called kernels, onto the GPU through CUDA streams, i.e., ordered queues for asynchronous execution. A kernel executes many lightweight GPU threads organized hierarchically into a grid of thread blocks, also known as Cooperative Thread Arrays (CTAs). Threads within a CTA cooperate through shared memory and synchronization, while CTAs are independent and can be scheduled in parallel across the GPU~\cite{cudaProgrammingGuide}.

The primary compute units of the GPU are SMs, analogous to CPU cores. Each SM provides the resources required to execute GPU threads, including registers, shared memory, caches, and arithmetic units. When a kernel is launched, its CTAs are distributed across the SMs, with each CTA assigned to a single SM until completion. Multiple CTAs may reside on the same SM, subject to resource availability. At a finer granularity, threads execute in warps, typically groups of 32 threads following the SIMT (Single Instruction, Multiple Threads) execution model~\cite{cudaProgrammingGuide}.

This organization makes the number of SMs assigned to a workload a direct proxy for its available GPU compute capacity. If a workload exposes enough independent CTAs, increasing the SM allocation can reduce completion time by enabling more parallel execution.

\subsection{Fundamental GPU Partitioning Primitives}

\subsubsection{Temporal and hardware-level sharing.} 
Introduced with NVIDIA's 2012 Kepler architecture, Hyper-Q provides multiple hardware work queues, enabling concurrent kernel execution. Building on this capability, MPS~\cite{MPS} merges CUDA contexts into a single execution context, allowing concurrent kernel execution across processes. Recent MPS versions additionally support soft partitioning through per-client SM limits and hard partitioning via static SM allocation. In contrast, MIG~\cite{MIG} provides hardware-enforced spatial partitioning by dividing a GPU into isolated instances with dedicated compute, memory, and bandwidth. While MPS enables fine-grained sharing, MIG offers stronger isolation but is limited to a small number of fixed-size partitions without runtime resizing. See~\cite{villarrubia2026comprehensive} for a comparison of MPS and MIG.

\subsubsection{Driver-Level dynamic sharing.} 
To overcome the limited flexibility of static partitioning, NVIDIA introduced GCs~\cite{GC}, a lightweight CUDA driver abstraction enabling dynamic SM-level partitioning. Indeed, Green Contexts expose fine-grained SM partitions that can be managed directly by the runtime, providing a more flexible substrate for dynamic resource sharing. GCs and MPS are complementary: GCs defines SM partitions, while MPS enables concurrent execution of multiple CUDA processes within each partition.

\subsection{Advanced GPU Sharing and Scheduling Frameworks}

\subsubsection{Static and coarse-grained approaches}
To improve utilization under isolation constraints, several works combine MIG and MPS mechanisms, including ParvaGPU~\cite{lee2024parvagpu}, which applies MPS within MIG instances to increase intra-partition utilization, MIGER~\cite{zhang2024miger}, which introduces hierarchical scheduling to jointly optimize MIG partition sizing and MPS sharing under QoS constraints, MISO~\cite{li2022miso}, and BOER~\cite{zhang2025boer}, which refine co-location strategies for heterogeneous workloads. KACE~\cite{han2024kace} further introduces prediction-based workload placement over MIG and MPS. These methods improve GPU utilization by combining spatial isolation with intra-partition concurrency, but remain fundamentally constrained by static or slowly adapting resource partitioning decisions.

\begin{figure*}[!h]
\centering
\includegraphics[width=\textwidth]{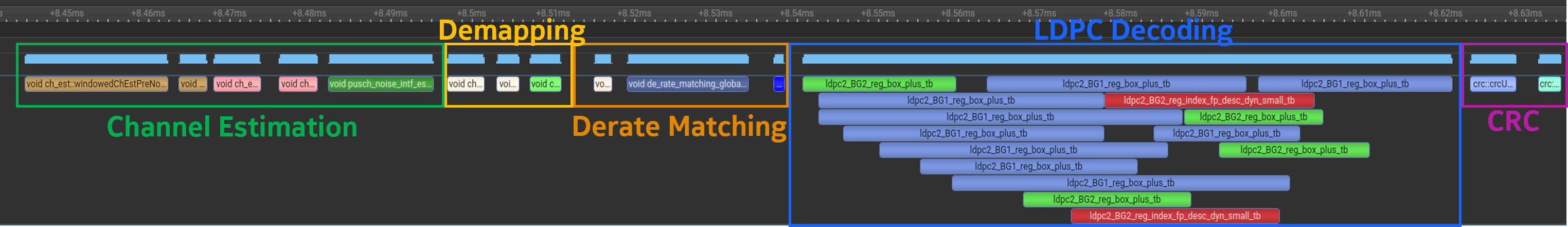}
\caption{GPU kernel execution trace of the PUSCH receive pipeline in NVIDIA Aerial cuPHY~\cite{nvidia_aerial_cuda_accelerated_ran} using NVIDIA Nsight Systems for profiling. }
\label{fig:ldpc}
\end{figure*}

\subsubsection{Dynamic and fine-grained approaches}

To enable fine-grained GPU sharing, several systems rely on runtime interception and kernel-level scheduling. MMK~\cite{jiang2026mmk} intercepts CUDA API calls and dynamically controls kernel execution based on runtime metrics such as SM utilization and memory bandwidth. Similarly, Draco~\cite{radig2025draco} leverages GCs and runtime monitoring to dynamically reallocate SM partitions between latency-critical inference and best-effort workloads. HAMI~\cite{HAMI} extends this approach to Kubernetes by exposing SM and memory partitioning through CUDA interception. Tally~\cite{zhao2025tally} further introduces a virtualization layer combining API interception with just-in-time (JIT) kernel transformation to support kernel slicing and opportunistic scheduling.

Overall, these approaches improve GPU sharing but often depend on complex CUDA interception, application restructuring, or kernel transformation. In contrast, ScaleRAN relies on a lightweight runtime library to enable fine-grained GPU sharing with minimal application modifications.


\subsection{GPU Resource Management in AI-RAN}

Existing AI-RAN studies primarily rely on hardware partitioning to isolate baseband processing and AI inference tasks. For example,~\cite{yet2026sla} uses MIG to co-locate 5G baseband functions and AI inference workloads. 
Beyond static partitioning,~\cite{shah2025proactive} proposes a reinforcement learning-based orchestrator that dynamically allocates MIG resources across RAN and AI workloads using KPI monitoring exposed via the Y1 O-RAN interface. Complementary studies, such as~\cite{barker2026six}, demonstrate the benefits of GPU acceleration for LDPC decoding, showing improved throughput and reduced latency, reinforcing the feasibility of GPU-centric AI-RAN deployments.

The recent AI-RAN Alliance white paper on the \emph{Platform and Infrastructure Orchestrator} highlights the need for an intelligent O-Cloud runtime to coordinate heterogeneous compute resources across RAN and AI workloads~\cite{airan_orchestrator2025}. It identifies dynamic resource allocation, proactive autoscaling, workload placement, and accelerator-aware orchestration as key capabilities. ScaleRAN addresses these requirements through a node-local GPU resource manager that combines fine-grained Green Context partitioning with reactive and predictive autoscaling to dynamically redistribute GPU resources between latency-critical RAN processing and co-located AI applications.

\section{RAN and AI Workloads}
\label{sec:workloads}

This section introduces the RAN workload characteristics that motivate dynamic GPU sharing with AI.

\begin{figure}[!hb]
\centering
\includegraphics[width=\columnwidth]{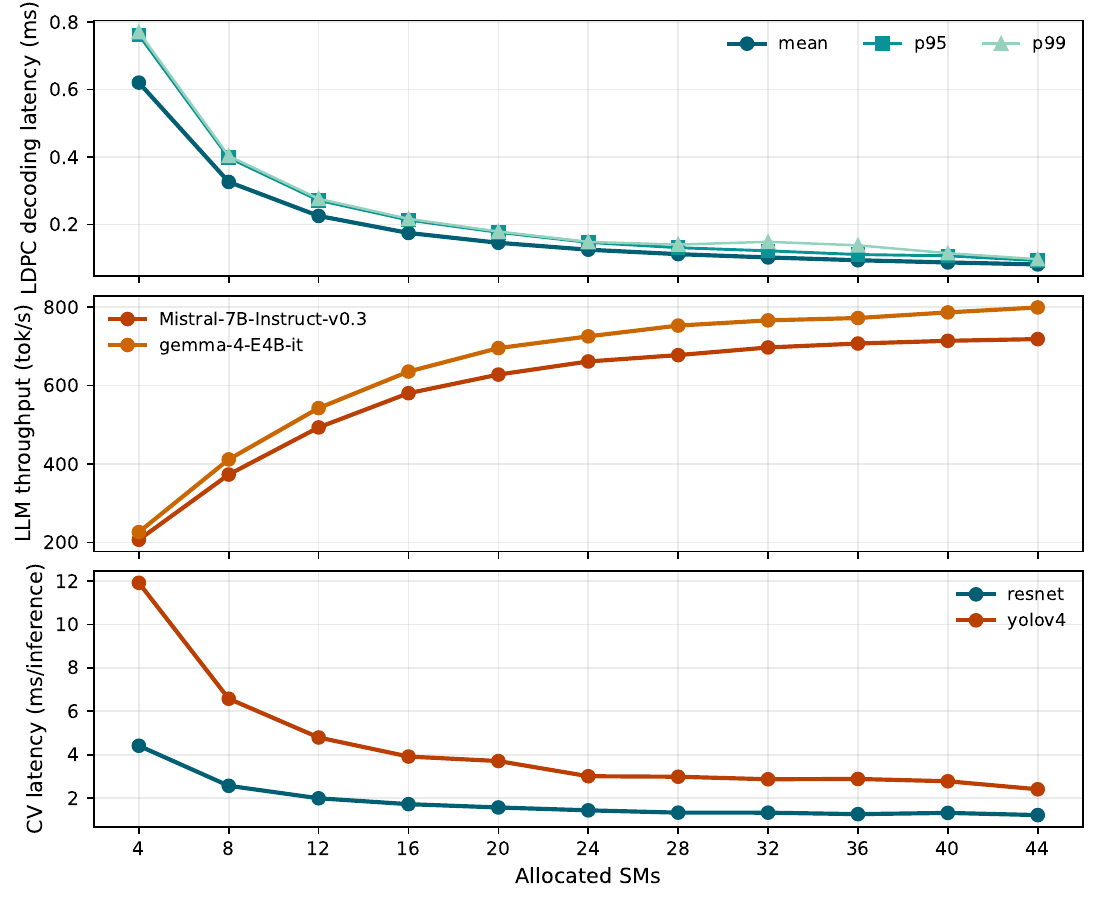}
\caption{Performance of RAN and AI workloads as a function of allocated  SMs on an NVIDIA DGX Spark (48 SMs in total). LDPC plot is averaged over 2000 slots (cuPHY~\cite{nvidia_aerial_cuda_accelerated_ran}). LLM curves report aggregate token generation throughput with a prompt batch size of 64 (prompts of 40-120 characters).}
\label{fig:sm-alloc}
\end{figure}

\subsection{RAN Workload: Predominant LDPC Decoding}
\label{sec:ldpc}

In 5G NR, Low-Density Parity-Check (LDPC)~\cite{tarver2021gpu} codes provide forward error correction for the high-throughput data channels (i.e., PDSCH/PUSCH). On the receive path, the decoder takes the bit estimates (log-likelihood ratios) produced by demodulation and reconstructs the transport-block bits. Fig.~\ref{fig:ldpc} shows a GPU kernel execution trace of the PUSCH receive pipeline in NVIDIA Aerial cuPHY~\cite{nvidia_aerial_cuda_accelerated_ran}, from channel estimation and demapping through de-rate matching, LDPC decoding, and CRC. LDPC decoding occupies by far the largest portion of the GPU execution timeline, making it the dominant computational component and the target for elastic GPU resource allocation. Its workload scales with the offered traffic (number of users, modulation and coding scheme, and allocated bandwidth), and all code blocks must be decoded within the slot duration (e.g., $0.5$,ms for numerology $\mu=1$).

LDPC decoding is particularly well suited to elastic GPU execution for two reasons. First, the code blocks within a slot are mutually independent and are dispatched as concurrent GPU kernels, allowing the workload to scale efficiently with the number of allocated SMs. Second, because LDPC both dominates the PUSCH computational cost and exhibits a regular, highly parallel structure, it provides a representative latency-critical RAN workload for studying GPU resource sharing. Consequently, our simulator and experiments focus on reproducing its GPU execution behavior to capture the dominant compute dynamics of the RAN pipeline (Sec.~\ref{sec:LDPCsim}).

Fig.~\ref{fig:sm-alloc} illustrates how LDPC decoding latency decreases as additional SMs are allocated, enabling more code blocks to execute concurrently. Once all available parallelism has been exploited, however, further increasing SMs provides little benefit, as decoding latency becomes bounded by the decoder's critical path. In practice, the number of SMs required to meet the slot deadline depends on the offered load, which determines the number of code blocks to decode. Since SMs are scarce resources, over-provisioning the RAN workload wastes GPU capacity that could otherwise be allocated to AI workloads, whereas under-provisioning risks violating RAN timing constraints.  Efficient GPU sharing therefore requires allocating only the SMs needed to satisfy the current RAN deadline, leaving the remaining resources for AI workloads.

\subsection{AI Workloads}

As illustrated in Fig.~\ref{fig:sm-alloc}, deep learning inference workloads, ranging from convolutional neural networks (CNNs) such as ResNet to transformer-based large language models (LLMs), exhibit decreasing inference latency as additional SMs are allocated. Their execution consists of a sequence of GPU kernels implementing tensor operations such as matrix multiplications, convolutions, and attention mechanisms. While compiler and runtime optimizations (e.g., TensorRT) fuse multiple operations into larger kernels to improve throughput, these long-running kernels reduce scheduling opportunities because GPU kernels are non-preemptive. Conversely, decomposing inference into smaller kernels slightly decreases execution efficiency but creates more scheduling points, enabling finer-grained scheduling by the GPU runtime. Similar to LDPC decoding, inference latency exhibits diminishing returns as the number of allocated SMs increases, with the exact scaling depending on the model architecture and input characteristics (e.g., batch size, input dimensions, or LLM context length). This elastic relationship between inference latency and SM allocation provides the basis for dynamically reallocating GPU resources between AI and RAN workloads.

\section{ScaleRAN Runtime Architecture}
\label{sec:autoscaler}

This section presents the design and implementation of the ScaleRAN runtime.

\subsection{Elastic SM Partitioning with Green Contexts}
\label{sec:gc}

ScaleRAN relies on  GCs as the low-level mechanism for enforcing elastic GPU sharing. 
Alongside the use of GCs, it  also uses MPS to enable concurrent CUDA execution across the RAN and AI processes in their respective SM partitions. 

In ScaleRAN, GPU resource sharing is controlled using a finite ladder of SM allocation profiles,
\[
\Pi=\{\pi_0,\pi_1,\ldots,\pi_{P_{\max}}\}.
\]
Each profile $\pi_p\in\Pi$ defines a pair of disjoint GC partitions,

\[
\pi_p=\left(S^{\mathrm{RAN}}_p,S^{\mathrm{AI}}_p\right),
\]

where $S^{\mathrm{RAN}}_p$ and $S^{\mathrm{AI}}_p$ denote the number of SMs allocated to the RAN and AI workloads, respectively. The profiles are ordered such that increasing the profile index $p$ allocates more SMs to the RAN workload and fewer SMs to the AI workload.

All profiles are created during initialization. At runtime, ScaleRAN adjusts the GPU allocation by switching between profiles, redirecting subsequent GPU work to the corresponding GC. Since CUDA kernels are non-preemptive, profile changes take effect only at scheduling boundaries, i.e., between kernel launches, without interrupting kernels already executing on the GPU. This approach avoids device reconfiguration, incurs minimal control overhead, and provides sufficient granularity for slot-level RAN adaptation. 

\subsection{Autoscaling Problem}
\label{sec:problem}

The objective of the autoscaler is to allocate the minimum GPU resources to the RAN workload while satisfying its real-time deadline. In general, the proposed framework is applicable to any latency-critical GPU-accelerated RAN function and to multiple co-located AI workloads. For the remainder of this paper, we consider the RAN workload to be the LDPC decoding task, which dominates the GPU execution time of the 5G uplink processing pipeline (see Sec.~\ref{sec:ldpc}), and assume a single AI workload co-located with the RAN function. The extension of the proposed architecture to multiple heterogeneous AI workloads is discussed in Sec.~\ref{sec:lib}.

The resource-allocation problem can be formulated as

\begin{align}
\min_{p_t \in \Pi} \quad & S^{\mathrm{RAN}}_{p_t}
\\
\text{s.t.}\quad
& \Pr\!\left(d_t(p_t)\leq D\right)\geq 1-\epsilon,
\end{align}

where $d_t(p_t)$ denotes the LDPC decoding latency obtained with profile $p_t$ at time $t$, $D$ is the slot processing deadline, and $\epsilon$ is the maximum acceptable deadline violation probability.

\subsection{Architecture}
\label{sec:archi}

\begin{figure}[t]
\centering
\includegraphics[width=\columnwidth]{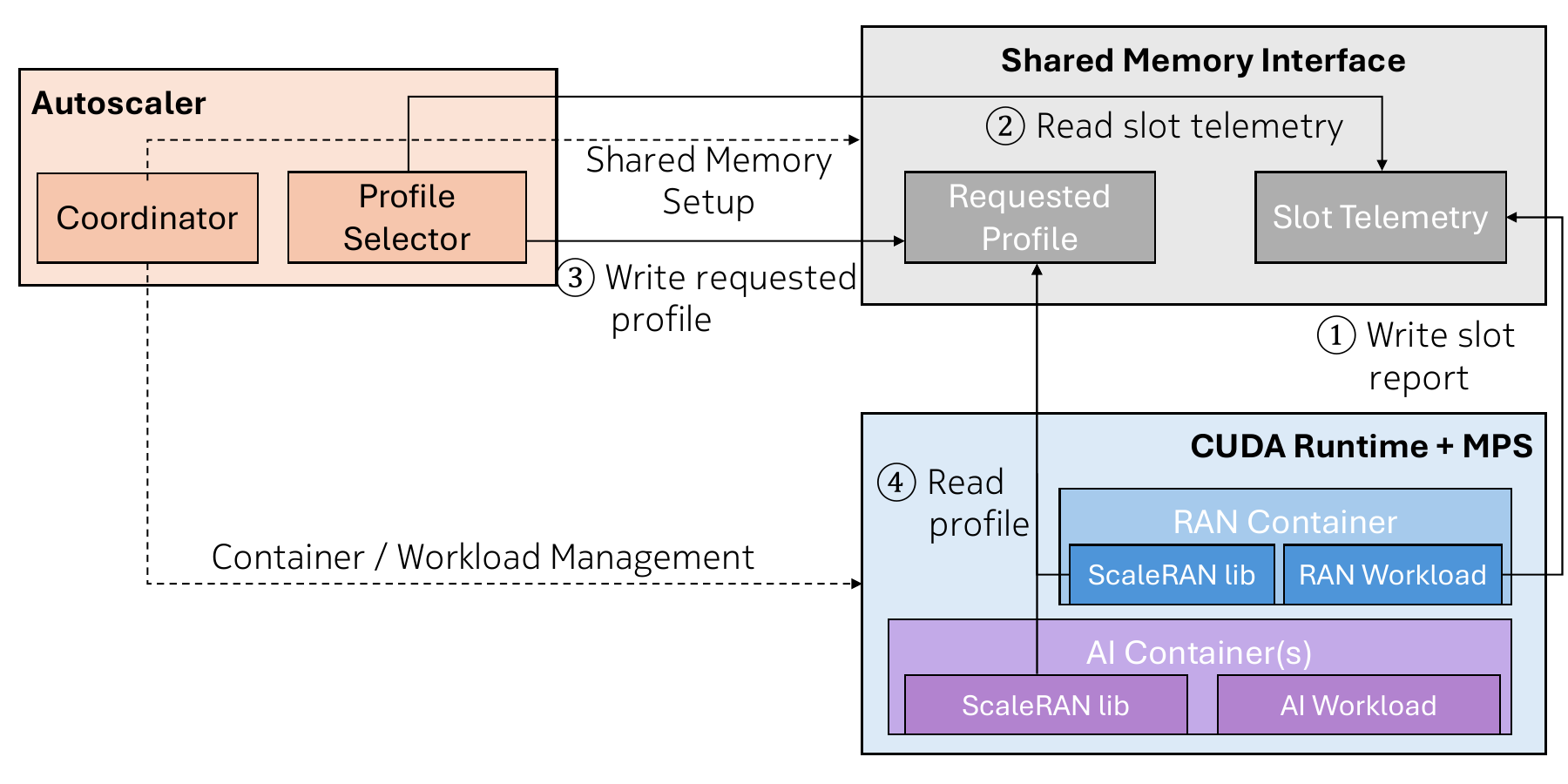}
\caption{ScaleRAN runtime architecture on a shared GPU. At initialization, the \emph{Coordinator} in the autoscaler sets up POSIX shared-memory control structures, creates the Green Context profile ladder (Sec.~\ref{sec:gc}) and deploys the RAN and AI containers. 
On the fast control path, the RAN workload writes per-slot telemetry (LDPC decoding latency and workload descriptor) to a ring buffer \textbf{(1)}, which the \emph{Profile Selector} in the autoscaler reads \textbf{(2)} to select the target profile from the predefined ladder. The selector writes the requested profile index to a shared control block \textbf{(3)}, which the ScaleRAN library in each container reads \textbf{(4)} to switch the active Green Context (and CUDA Streams) before subsequent GPU work is launched.}
\label{fig:scaleran}
\end{figure}

ScaleRAN follows a split-process architecture composed of a RAN worker, an AI worker, an autoscaler, and a lightweight coordinator, as illustrated in Fig.~\ref{fig:scaleran}. 
The RAN worker executes slot-level processing under strict timing constraints, while the AI worker executes best-effort inference and the autoscaler dynamically adjusts SM partitions between the two workloads.

During initialization, the \emph{coordinator} module of the autoscaler constructs the GC profile ladder $\Pi$, creates  shared-memory control structures, and launches the worker processes. Since every profile deterministically defines SM partitions, each worker independently reconstructs the same profile ladder and selects its own partition, avoiding the need for a centralized CUDA context manager.

The interaction between workers and the autoscaler consists of two communication flows implemented through shared memory. First, the RAN worker continuously publishes slot reports containing the measured LDPC decoding latency together with optional workload descriptors, such as the number of allocated Physical Resource Blocks (PRBs), active User Equipments (UEs), Modulation and Coding Scheme (MCS) statistics, code blocks, CTAs, and base graph.
Then, the \emph{Profile Selector} module of the autoscaler consumes this telemetry and executes the selected scaling policy to determine the appropriate GPU allocation. 
The selected profile index $p$ is written to shared memory and read by both workers. When RAN and AI worker get an updated profile, they execute subsequent kernels in their updated GC. As mentioned above, CUDA kernels are non-preemptive and profile updates can only become effective at scheduling boundaries.

\subsection{Implementation}
\label{sec:lib}

Runtime coordination between the workers and the Profile Selector relies on two POSIX shared-memory regions: 
a lock-free single-producer/single-consumer ring buffer carrying slot telemetry from the RAN worker to the Profile Selector, and a compact control block storing the selected profile, synchronization state, and worker status. These shared-memory regions are hosted by the RAN container and are accessed through the ScaleRAN library described below, which abstracts the communication primitives from both the RAN and AI workers. This design avoids inter-process message passing on the critical control path while keeping the runtime overhead of autoscaling low.

ScaleRAN targets GPU-enabled edge platforms where RAN and AI functions execute in  containers sharing the same GPU. The coordinator deploys and manages both containers through the Docker Engine API. The AI container joins the IPC namespace of the RAN container, allowing both containers to access the same shared-memory control structures. 
Both containers connect to the same MPS server, allowing RAN and AI kernels to execute concurrently on their respective GC partitions rather than time-slicing the GPU.

Applications integrate with ScaleRAN through a lightweight \emph{ScaleRAN library} that attaches to the shared-memory control block, reconstructs the GC profile ladder, and creates one CUDA stream per profile. Integrating a new workload requires only wrapping GPU-related calls with the stream selector provided by the library. At each AI inference or LDPC decoding iteration, the library checks whether the autoscaler has selected a new profile and, if necessary, synchronizes the previous stream, activates the CUDA stream associated with the new GC, and acknowledges the profile change through shared memory. This cooperative approach requires only minor modifications to applications while remaining compatible with  libraries such as PyTorch, TensorRT, ONNX Runtime, and Hugging Face Transformers for AI applications. Fig.~\ref{fig:lib-example} illustrates the integration on a minimal inference loop.

\begin{figure}[t]
\begin{lstlisting}[style=scaleran, basicstyle=\ttfamily\scriptsize]
import torch
(*@\color{scaleranblue} from scaleran import ProfileManager@*)

model = load_model().cuda().eval()

(*@\color{scaleranblue}with ProfileManager() as pm:@*)
  for x in inputs:
    (*@\color{scaleranblue} with pm.synchronize():@*)
      y = model(x)
\end{lstlisting}
\caption{Injecting a ScaleRAN-managed AI workload. Code integration reduces to attaching to the autoscaler once (\texttt{ProfileManager}) and wrapping each inference in the Green Context stream returned by \texttt{pm.synchronize()}. Statements in blue are the only  required to integrate an existing workload.}
\label{fig:lib-example}
\end{figure}

\subsection{Discussion} 
\label{sec:discussion}

The library-based interface is intrusive, since it requires minor application modifications, and its switching granularity is limited to synchronization points rather than individual kernel launches. An alternative would be a transparent CUDA interception layer, which would eliminate application changes and enable profile switches at kernel granularity. However, such an approach would require intercepting a large number of CUDA API calls and migrating objects across GCs. For modern AI frameworks, this is both complex and fragile, motivating our simpler and more robust cooperative design.

Although the current prototype targets one RAN worker and one AI worker, the same architecture naturally extends to a larger number of co-located workloads. In this case, the profile index published by the autoscaler would identify a complete GPU partitioning across all workers rather than a single RAN/AI split. Each worker could independently reconstruct the GC associated with its own partition for the selected profile, while the synchronization mechanism remains unchanged. Because the number of possible profile combinations grows rapidly with the number of workloads, defining profiles at the granularity of worker classes rather than individual workers provides a more scalable design while preserving the simplicity of the control plane.

\subsection{O-RAN Integration}
\label{sec:oran}

In the O-RAN architecture, radio access network functions are disaggregated into the O-RU (Radio Unit), O-DU (Distributed Unit), and O-CU (Central Unit). These virtualized RAN functions execute on a shared cloud infrastructure, referred to as the O-Cloud, which provides the compute, storage, networking, and hardware accelerator resources required to host RAN and AI workloads. The O-RAN architecture further separates real-time control from management: the Near-Real-Time RAN Intelligent Controller (Near-RT RIC) hosts xApps for radio optimization at timescales above approximately 10 ms, while the Service Management and Orchestration (SMO) framework performs network-wide orchestration and policy management over longer timescales. In NVIDIA Aerial~\cite{nvidia_aerial_cuda_accelerated_ran}, the O-DU is implemented using the cuMAC and cuPHY software libraries, which provide GPU-accelerated implementations of the MAC (Layer 2) and PHY (Layer 1) functions, respectively. The cuMAC scheduler determines which user equipments (UEs) are served in each slot, together with their allocated PRBs, MCS, hybrid automatic repeat request (HARQ) transmissions, and multi-user scheduling decisions. These scheduling decisions define the transport and code blocks processed by cuPHY, including the uplink LDPC decoder, thereby shaping both the average GPU workload and its slot-level burstiness and parallelism. Consequently, scheduler decisions have a direct impact on the GPU resources required to satisfy RAN timing constraints.

\begin{figure}[t]
\centering
\includegraphics[width=0.8\columnwidth]{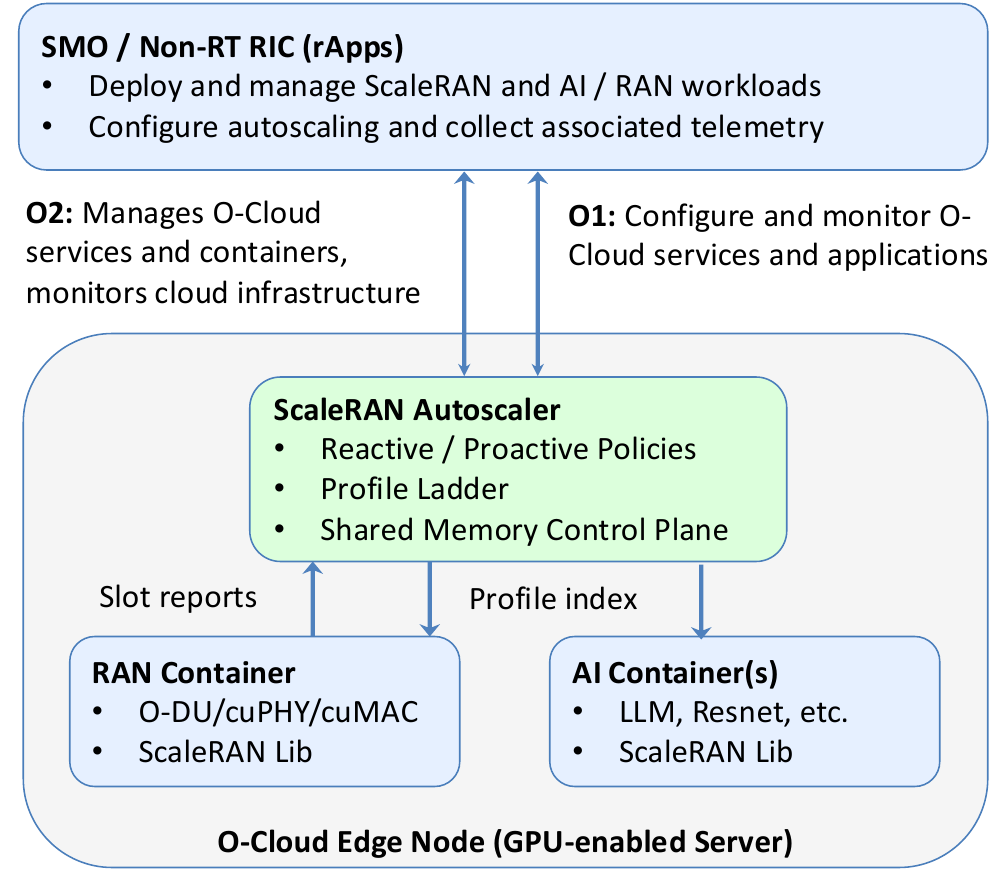}
\caption{Integration of ScaleRAN within the O-RAN architecture. ScaleRAN operates as a node-local O-Cloud service that performs slot-level GPU resource allocation. The SMO configures and monitors the autoscaler through O1/O2 interfaces, while Near-RT RIC applications provide optional policy guidance without participating in the real-time GPU control loop.}
\label{fig:oran-architecture}
\end{figure}

Fig.~\ref{fig:oran-architecture} illustrates how ScaleRAN integrates into the O-RAN architecture. ScaleRAN is deployed as a node-local O-Cloud service co-located with the O-DU and AI workload containers, rather than as a Near-RT RIC xApp. The slot-level autoscaling loop requires direct access to LDPC timing measurements together with low-latency control of GC profile switches, making local execution essential. Consequently, the real-time GPU control loop relies exclusively on shared-memory communication within the edge node, while the O-RAN management interfaces are reserved for orchestration, configuration, and monitoring.

Long-timescale optimization can be implemented through one or more ScaleRAN rApps executing within the SMO framework. These rApps can select the active autoscaling policy, deploy predictive models, configure controller parameters, or coordinate GPU resource management across edge nodes. The O2 interface deploys and manages the ScaleRAN service together with the RAN and AI containers, while monitoring GPU inventory and accelerator capabilities. The O1 interface configures runtime policies parameters, deadline targets, and resource limits—and collects telemetry such as the active profile, LDPC decoding latency, GPU utilization, AI throughput, profile-switch statistics, and deadline violations. This information enables operators and ScaleRAN rApps to assess SLA compliance, optimize resource utilization, and refine autoscaling policies.

As part of O-RAN WG6~\cite{oran_wg6} and its ongoing subtask \emph{O-Cloud Enhancement for AI Applications}, led by China Mobile, O-RAN is analyzing the impact of AI applications on the O-Cloud platform. The current work, still at an early draft stage, focuses on identifying requirements for GPU sharing and AI application offloading. In particular, it considers enhancements to the Accelerator Abstraction Layer (AAL) and the O2 interface to support orchestration of GPU-enabled and other heterogeneous compute nodes, including resource discovery, topology, monitoring, and fault management. AI-driven workload placement, elastic scaling, and performance optimization have been identified as key directions. The O-RAN integration presented in this paper demonstrates how ScaleRAN addresses these emerging requirements by operating as a node-local O-Cloud runtime for elastic GPU sharing between co-located RAN and AI workloads, while leveraging the O1 and O2 interfaces for configuration, orchestration, and monitoring.

\section{Autoscaling Policies}
\label{sec:policies}

ScaleRAN supports multiple autoscaling policies.
In this work, we consider two complementary approaches: a reactive controller inspired by  Kubernetes (Sec.~\ref{sec:k8s-controller}) and a proactive prediction-based controller (Sec.~\ref{sec:proactive}).

\subsection{Kubernetes-Inspired Policy}
\label{sec:k8s-controller}

The Kubernetes Horizontal Pod Autoscaler (HPA)~\cite{kubernetes_hpa} is one of the most widely deployed reactive autoscaling mechanisms in cloud computing and therefore serves as a natural baseline for comparison. The HPA adjusts the number of application replicas according to the ratio between an observed resource metric and a target value. Inspired by this principle, we consider a Kubernetes-inspired controller that adapts the RAN GPU allocation proportionally to the ratio between the measured LDPC decoding latency and the target deadline. Although HPA operates on second timescales, its control principle readily extends to slot-level GPU resource management by replacing replicas with SM allocations and utilization with LDPC decoding latency.

Let $d_{t-1}$ denote the decoding latency measured during the previous slot, $D$ the target decoding deadline, and $S^{\mathrm{RAN}}_{p_{t-1}}$ the current number of SMs allocated to the RAN workload. Following the proportional scaling rule of the HPA, the desired RAN allocation is computed as

\begin{equation}
\widehat{S}^{\mathrm{RAN}}_t
=
S^{\mathrm{RAN}}_{p_{t-1}}
\frac{d_{t-1}}{D}
\label{eq:k8s-desired-sms}
\end{equation}

The desired allocation is then mapped to the smallest profile providing at least the required number of SMs,

\begin{equation}
\widehat{p}_t
=
\min\!\left(
P_{\max},
\max\!\left(
0,
\min
\left\{
p\in\Pi
\;\middle|\;
S^{\mathrm{RAN}}_p
\ge
\widehat{S}^{\mathrm{RAN}}_t
\right\}
\right)
\right)
\end{equation}

Although the Kubernetes HPA  incorporates a \emph{deadband} to avoid frequent scaling actions, this mechanism is omitted since the controller must react immediately to increases in decoding latency to minimize RAN deadline violations.

To protect the RAN deadline, upscaling decisions are applied immediately,

\begin{equation}
p_t
=
\widehat{p}_t,
\qquad
\text{if }
\widehat{p}_t > p_{t-1}
\label{eq:k8s-upscale}
\end{equation}

Conversely, downscaling decisions are delayed over a stabilization window of $H$ slots. This mechanism prevents transient reductions in RAN load from prematurely releasing GPU resources.  A lower profile is applied only after no higher allocation has been requested during the stabilization window. Let

\begin{equation}
\mathcal{P}_t
=
\left\{
\widehat{p}_{\max(0,t-H+1)},
\ldots,
\widehat{p}_t
\right\}
\label{eq:k8s-window}
\end{equation}

denote the candidate profiles computed during the stabilization window. When the current candidate requests fewer resources, the controller selects the largest profile requested during the window,

\begin{equation}
p_t
=
\max_{q \in \mathcal{P}_t} q,
\qquad
\text{if }
\widehat{p}_t < p_{t-1}
\label{eq:k8s-downscale}
\end{equation}

\begin{table}[t]
\centering
\caption{Features used by the decoding-latency predictor.}
\label{tab:ml-features}
\scriptsize
\begin{tabular}{|p{1.6cm}|p{4.4cm}|p{1cm}|}
\hline
\textbf{Feature} & \textbf{Description} & \textbf{Type} \\
\hline
Allocated PRBs & Number of PRBs allocated during the slot & Integer \\
MCS mean & Mean of MCS across scheduled UEs & Numerical \\
Layers & Total number of MIMO layers & Integer \\
Coded bits & Total number of coded bits to be processed  & Integer \\
Base graph & LDPC base graph (BG1 or BG2) & Categorical \\
SM profile $p$ & Current SM allocation profile & Categorical \\
\hline
\end{tabular}
\end{table}

\subsection{Proactive Policy}
\label{sec:proactive}

The proactive controller instead predicts the decoding latency of the next slot from radio descriptors
$\mathbf{x}_t$ (e.g., allocated PRBs, active UEs, and MCS statistics),

\begin{equation}
\hat d_t = g(\mathbf{x}_t),
\end{equation}

and selects the smallest profile satisfying

\begin{equation}
p_t=
\min
\left\{
p\in\Pi
\;\middle|\;
\hat d_t(p)\leq D
\right\}.
\end{equation}

To improve the safety and minimize the risk of missing RAN deadlines, the proactive policy is based on a  machine-learning model that predicts a high conditional quantile of the decoding latency rather than its expected value. 
Specifically, it estimates the conditional quantile at level $\tau$, where $\tau = 1-\epsilon$,

\begin{equation}
\widehat{d}^{(\tau)}_t(p)
=
g_{\tau}(\mathbf{x}_t,p),
\end{equation}

where $\mathbf{x}_t$ denotes the radio descriptors (e.g., allocated PRBs, active UEs, and MCS statistics), $p$ is the candidate RAN profile, and $g_{\tau}(\cdot)$ is trained to estimate the conditional decoding-latency quantile at level $\tau$ (e.g., 95\% quantile).

The set of input features (i.e., $\mathbf{x}_t$, $p$) used by the prediction model is summarized in Table~\ref{tab:ml-features}. These features characterize the instantaneous decoding workload independently of any traffic scenario, allowing the predictor to learn a generic mapping from radio workload descriptors to decoding latency.

The model is trained using the Pinball (quantile) loss,

\begin{equation}
\mathcal{L}_{\tau}(d,\hat{d})=
\begin{cases}
\tau(d-\hat{d}), & d\geq\hat{d},\\[2mm]
(1-\tau)(\hat{d}-d), & d<\hat{d},
\end{cases}
\end{equation}

where $\tau\in(0,1)$ denotes the target quantile level.

The predictor is implemented as a gradient-boosted decision tree (GBDT) model using LightGBM. It takes as input the radio and execution descriptors summarized in Table~\ref{tab:ml-features}. This class of models is particularly suitable because it captures complex nonlinear relationships between the decoding workload, resource allocation, and execution time while providing deterministic, low-latency inference suitable for slot-level execution. For deployment, we leverage lleaves~\cite{Boehm_lleaves}, which compiles the trained LightGBM model into LLVM-generated native code. This approach reduces inference overhead while preserving identical prediction results.

\textit{Conformal calibration.}
Although the quantile predictor directly estimates an upper bound on the decoding latency, its predictions may not achieve the desired empirical coverage. To address this issue, we apply one-sided split conformal prediction, which provides statistically valid upper bounds without modifying the underlying regression model. The available samples are partitioned into disjoint training, calibration, and test sets. The predictor $g_{\tau}$ is first trained on the training set to estimate the conditional quantile. Then, for each sample $i$ in the calibration set
$\mathcal{D}_{\mathrm{cal}}$, we compute the one-sided
nonconformity score

\begin{equation}
r_i
=
\max
\left(
0,\,
d_i-\widehat{d}^{(\tau)}_i
\right),
\label{eq:conformal-score}
\end{equation}

Positive scores correspond to underestimation of the decoding
time, the relevant direction for deadline
protection.

Let $n_{\mathrm{cal}}
=
\left|\mathcal{D}_{\mathrm{cal}}\right|$ denote the number of calibration samples, and let $r_{(1)}
\leq
r_{(2)}
\leq
\cdots
\leq
r_{(n_{\mathrm{cal}})}$
be the calibration scores sorted in nondecreasing order. Given a target conformal miscoverage probability $\alpha$, the
conformal correction is defined as $q_{\alpha} = r_{(k)}$ with $k =
\min
\left(
n_{\mathrm{cal}},
\left\lceil
(n_{\mathrm{cal}}+1)(1-\alpha)
\right\rceil
\right)$. For a candidate profile $p$, the calibrated upper estimate of
the decoding latency is 

\begin{equation}
U_t(p)
=
\widehat{d}^{(\tau)}_t(p)
+
q_{\alpha}.
\label{eq:calibrated-upper-bound}
\end{equation}

The conformal correction guarantees a coverage of at least $1-\alpha$, limiting the probability of underestimating the decoding latency.

\begin{figure}[t]
\centering
\includegraphics[width=\columnwidth]{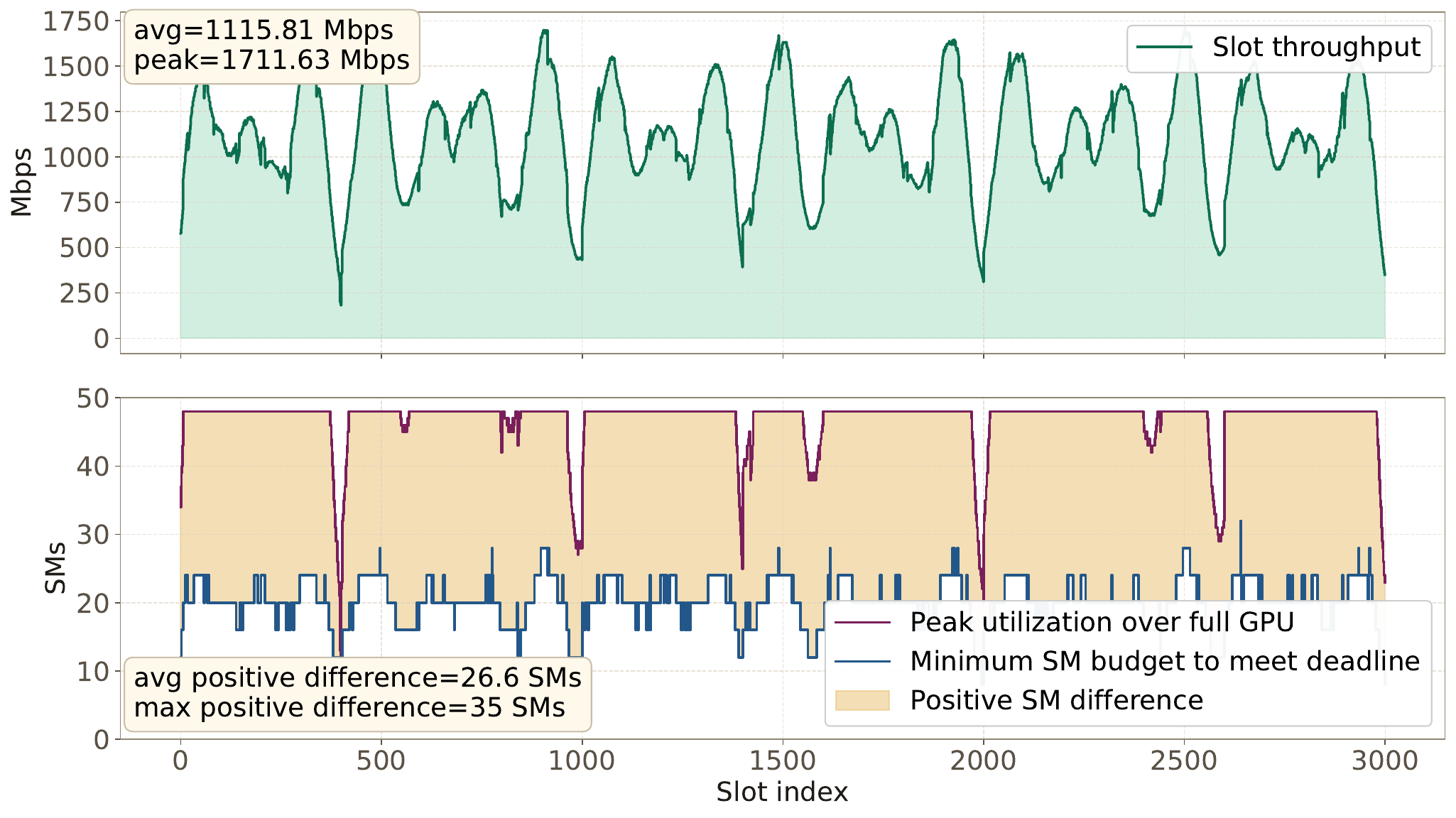}
\caption{Per-slot traffic and LDPC GPU resource demand under dynamic traffic conditions with 7 UEs. The upper panel shows the aggregate uplink throughput in Mbit/s. The lower panel compares the peak number of decoder SMs measured on the uncapped NVIDIA DGX Spark GPU (48 SMs) with the minimum, optimal 
SM allocation required to satisfy the 0.5 ms slot deadline. The shaded region indicates the positive difference between the timed peak demand and the optimal allocation, representing potential over-allocation.}
\label{fig:traffic-scenario}
\end{figure}

\section{Performance Evaluation}
\label{sec:eval}

This section evaluates the proposed ScaleRAN framework and compares the performance of the autoscaling policies.

\subsection{LDPCSim: RAN Workload Generation} \label{sec:LDPCsim} 

To evaluate ScaleRAN under representative GPU-enabled RAN workloads, we developed \emph{LDPCSim}, a Python-based workload generator built on NVIDIA Aerial CUDA-Accelerated RAN (ACAR)~\cite{nvidia_aerial_cuda_accelerated_ran}. LDPCSim provides a configurable interface to generate slot workloads for Aerial's cuPHY decoder, submitting them to the GPU, and collecting performance measurements.

The simulator converts a 5G NR scheduling configuration into a sequence of slot-level decoding workloads. For each scheduled user equipment (UE), it derives the transport-block size and LDPC parameters from radio settings including the allocated physical resource blocks (PRBs), modulation and coding scheme (MCS), and number of MIMO layers. PyAerial utilities perform the corresponding 3GPP-compliant transport-block sizing and LDPC segmentation calculations. Consequently, the decoder workload is determined by valid physical-layer configurations rather than by independently selected synthetic LDPC parameters.

The simulator supports static and time-varying multi-UE scenario configurations. In dynamic scenarios, each UE has a configurable traffic pattern, such as constant-bit-rate, sinusoidal and triangular waves, or periodic-burst traffic. A slot-level allocator distributes the shared PRB budget among active UEs. Changes in traffic arrivals and queue occupancy therefore lead to varying transport-block sizes, code-block counts, and GPU decoding demand. Fig.~\ref{fig:traffic-scenario} illustrates this variation through the aggregate traffic demand and the minimum SM allocation required to meet the 0.5 ms slot deadline.

For timed execution, LDPCSim groups compatible code blocks and submits them through cuPHY's batched decoding API. Independent groups are dispatched asynchronously on multiple CUDA streams and synchronized at slot completion. For each slot, the simulator records the decoding latency, measured on the CPU from kernel launch to completion. Unlike the GPU-side execution time, this measurement includes runtime overheads such as kernel launches and host-device synchronization, making it representative of the end-to-end slot processing time. The recorded decoding latency is used both as the RAN telemetry for the autoscaler and to evaluate ScaleRAN's deadline compliance.

\begin{figure}[t]
\centering
\includegraphics[width=0.85\columnwidth]{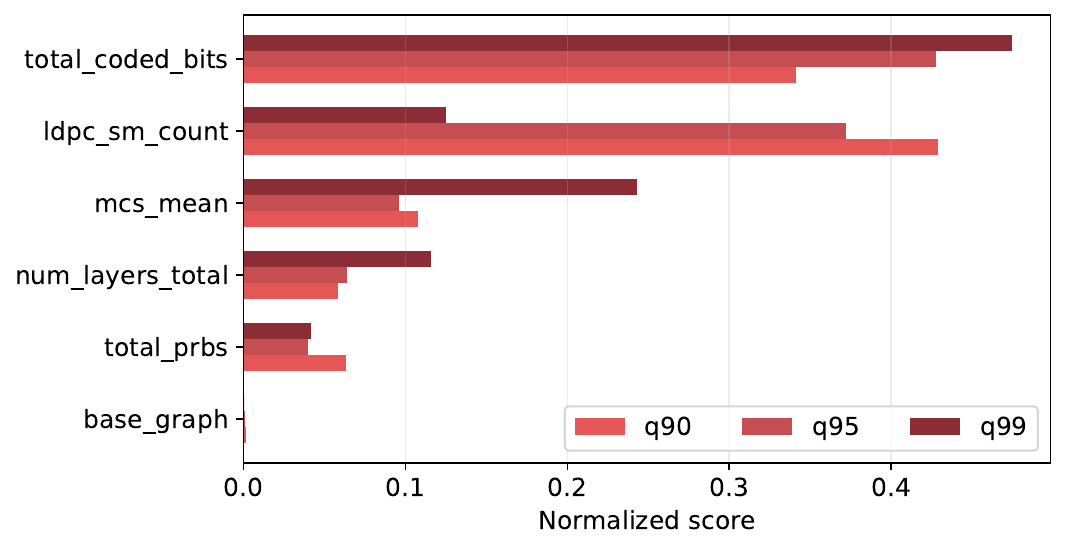}
\caption{Feature importance (gain) of the LightGBM quantile regression models trained to predict the P90, P95, and P99 decoding-latency quantiles.}
\label{fig:feature_importance}
\end{figure}

\subsection{Evaluation Setup}
\label{sec:methodo}

All experiments are conducted on an NVIDIA DGX Spark platform equipped with 48 SMs, using cuPHY for 5G uplink LDPC decoding with numerology $\mu=1$ and a slot deadline of $D=0.5$~ms. To evaluate the runtime under a conservative decoder configuration representative of low-SNR operation, we disable early termination and set the maximum number of LDPC iterations to 35. ScaleRAN uses an 11-profile Green Context ladder that assigns $S^{\mathrm{RAN}} \in \{4,8,\ldots,44\}$ SMs to the RAN workload in increments of $4$, while ensuring that both the RAN and AI workloads are allocated at least $4$ SMs.

To avoid leakage between model development and policy evaluation, we generate two disjoint datasets of traffic traces: a training dataset and an evaluation dataset. The training dataset consists of $105$ traces composed of $3000$ slots each spanning a wide range of radio configurations, including single-UE PRB sweeps across MCS and MIMO-layer configurations together with dense multi-UE scenarios. For every slot of every trace, LDPCSim measures the decoding latency under each profile of the GC ladder, yielding $3,465,000$ supervised samples that map the slot-level radio descriptors and candidate RAN allocation to the corresponding decoding latency. These samples are used exclusively to train the LightGBM predictor.

The evaluation dataset consists of $8$ previously unseen slots traffic traces made of $10,000$ slots each with a mix of heavy sinusoidal and bursty traffic patterns. For each evaluation trace, the first $500$ slots are reserved for conformal calibration, while the remaining $9500$ slots are used for policy evaluation. The calibration slots are used exclusively to estimate the conformal correction and are never used to evaluate the controllers. In practice, this calibration step could be periodically repeated online to adapt the predictor to changing traffic conditions. The quantile predictor is trained for a target quantile level $\tau$. Unless stated otherwise, when conformal calibration is applied, the target miscoverage probability is set to $\alpha = 1 - \tau$.

\subsection{Decoding-latency Predictor}

\begin{figure}[t]
\centering
\includegraphics[width=0.85\columnwidth]{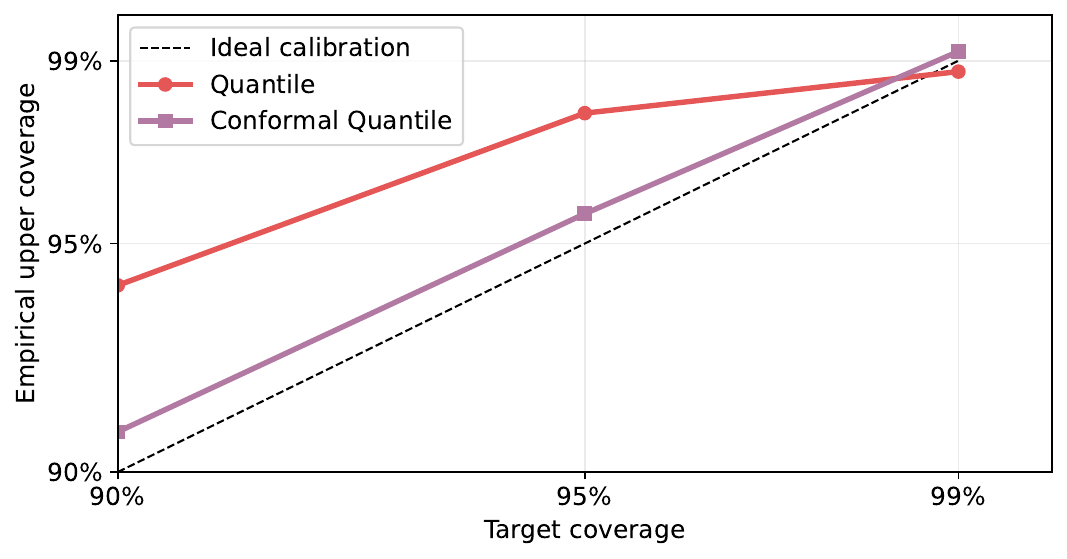}
\caption{Reliability diagram comparing empirical and target coverage before and after conformal calibration. The diagonal indicates perfect calibration.}
\label{fig:coverage}
\end{figure}

We first evaluate the decoding-latency predictor independently of the autoscaling policy. Fig.~\ref{fig:feature_importance} reports the feature importance of the LightGBM  models trained for the P90, P95, and P99 quantiles. The feature importance analysis indicates that decoding latency is primarily driven by the offered RAN workload. The total number of coded bits and the decoder SM count consistently dominate the predictions across all quantiles, whereas radio configuration parameters contribute less and the LDPC base graph has negligible importance. Next, we evaluate the effectiveness of conformal calibration. Fig.~\ref{fig:coverage} presents a reliability diagram comparing the empirical coverage obtained before and after calibration over multiple target quantiles. An ideal predictor lies on the diagonal, indicating that the empirical coverage matches the target coverage. The results show that conformal calibration restores the desired coverage while limiting the probability of underestimating the decoding latency. Finally, we assess the deployment overhead of the predictor. Using the LLVM-compiled lleaves implementation, a LightGBM batched inference requires only 60~$\mu$s, well below the 0.5~ms slot duration, demonstrating that it can execute online within the slot budget.

\subsection{Autoscaling Results}

\begin{figure*}[!t]
\centering
\includegraphics[width=\textwidth]{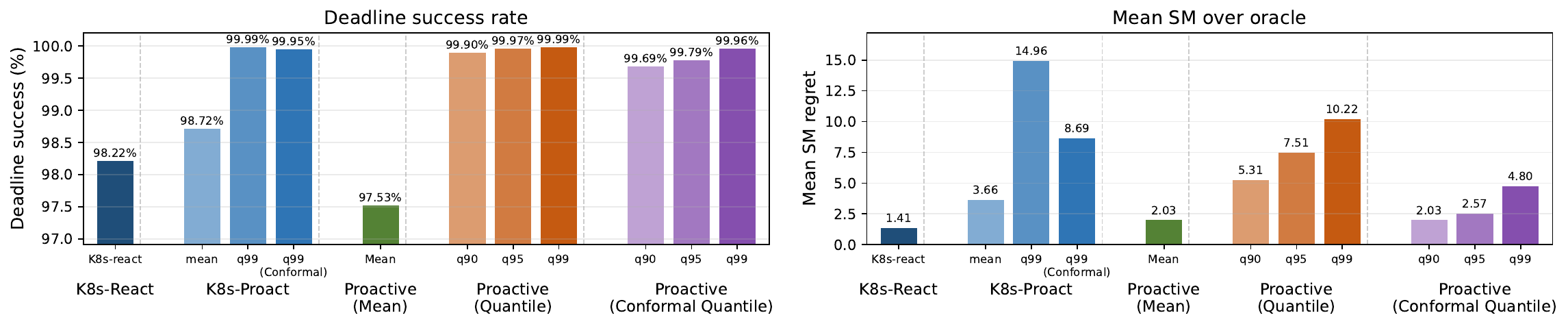}
\caption{Deadline success rate (left) and mean SM regret (right) for the evaluated autoscaling policies. SM regret is measured relative to the optimal allocation.}
\label{fig:missrate_regret}
\end{figure*}

We compare three autoscaling policies. The first, denoted K8S-React, is a reactive controller inspired by Kubernetes (Sec.~\ref{sec:k8s-controller}), which adjusts the RAN SM allocation using proportional control with a stabilization window based on the decoding latency measured in the previous slot (Sec.~\ref{sec:proactive}). The second, K8S-Proact, replaces the measured decoding latency with a prediction while preserving the same control law, thereby isolating the benefit of proactive execution. The third, Proactive, directly selects the RAN SM allocation from the predicted decoding latency without proportional control. For the predictive policies (K8S-Proact and Proactive), we evaluate three prediction variants: Mean, which predicts the expected decoding latency; Quantile, which predicts the P90, P95, or P99 decoding-latency quantile; and Conformal Quantile, which applies conformal calibration to the predicted quantiles. 

Fig.~\ref{fig:missrate_regret} illustrates the trade-off between deadline reliability and GPU overprovisioning across the evaluated autoscaling policies. Compared with the reactive Kubernetes-inspired controller, proactive prediction substantially improves the deadline success rate. Using higher prediction quantiles further increases reliability, approaching 100\% deadline success, at the cost of a larger SM regret. The proportional controller and stabilization window inherited from Kubernetes provide little benefit in our setting. Because decoding-latency predictions are highly accurate and LDPC decoding latency decreases nonlinearly with the number of allocated SMs (Fig.~\ref{fig:sm-alloc}), directly selecting the predicted SM allocation is more effective than relying on proportional control and a stabilization window. This is illustrated by the conformal P99 predictor, where K8S-Proact and Proactive achieve nearly identical deadline success rates (99.95\% versus 99.96\%), while Proactive reduces the mean SM regret by 45\%, from 8.69 to 4.80.

\subsection{AI and RAN interferences}

To characterize the interference between AI and RAN workloads, we evaluate the Proactive autoscaling policy based on conformal P99 prediction under a range of Deep Neural Network (DNN) architectures. The AI workloads include ResNet-18 and ResNet-152, together with dense Multi-Layer Perceptrons (MLPs) of varying depth (3, 10, and 40 layers) and hidden-layer width (1024 and 2048 neurons). These models span a broad range of execution characteristics, enabling us to investigate how they affect RAN deadline compliance.

\begin{figure}[t]
\centering
\includegraphics[width=\columnwidth]{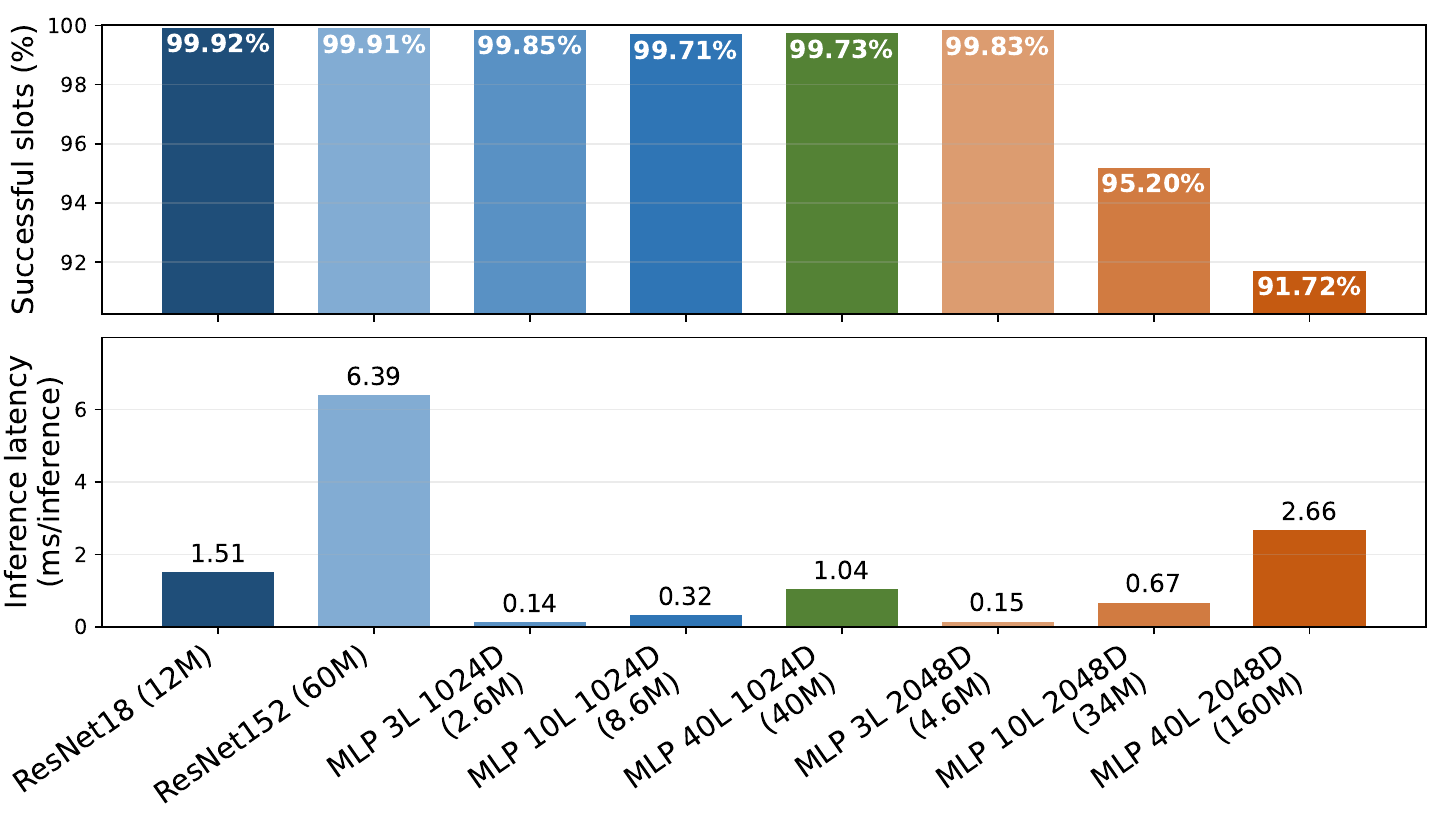}
\caption{Performance of Proactive (conformal P99) for different AI workloads. Top: RAN slots meeting the decoding deadline. Bottom: AI inference latency. Workloads include ResNet-18, ResNet-152, and MLPs with varying depth and hidden-layer width; parentheses indicate model size (parameters).}
\label{fig:interferences}
\end{figure}

Fig.~\ref{fig:interferences} shows that AI-RAN interference is governed by two complementary factors. The first is temporal interference, determined by the duration of AI kernel execution relative to the RAN slot duration. Since CUDA kernels are non-preemptive, kernels extending beyond a slot delay a new SM allocation after a profile change. The second is spatial interference, determined by the SM footprint of the AI kernels. Workloads that fully occupy their allocated SMs are more likely to interfere with subsequent RAN slots when the autoscaler changes the SM allocation. 

The results illustrate these two effects. Although ResNet-152 exhibits the longest inference latency (6.39 ms), it achieves nearly the same deadline success rate as ResNet-18, suggesting that its kernels do not fully occupy the allocated SMs and therefore leave sufficient opportunities for LDPC decoding. In contrast, the MLPs, particularly the widest ones with 2048 hidden units, exhibit substantially lower success rates, suggesting that their kernels occupy most or all of the allocated SMs and execute long enough to delay profile transitions. This is further illustrated by the 40-layer MLPs: although both exceed the 0.5 ms slot duration, only the 2048-hidden-unit model causes severe interference, confirming that SM footprint, rather than inference time alone, drives interference. Likewise, the 3-layer MLP with 2048 hidden units maintains a high deadline success rate despite its large SM footprint because its inference time is below the 0.5 ms RAN slot duration, allowing profile transitions to stay synchronized with RAN slot boundaries. These results show that neither end-to-end inference latency nor SM footprint alone predicts AI-RAN interference. Instead, interference is determined by the combination of SM footprint and kernel execution duration relative to the ScaleRAN control timescale. When kernel launch boundaries occur sufficiently frequently, profile transitions remain synchronized with RAN slot boundaries, protecting the RAN workload.

\section{Conclusion and Perspectives}
\label{sec:conclusion}

This paper presented ScaleRAN, a runtime for elastic GPU resource management in AI-RAN. By combining NVIDIA Green Context-based SM partitioning with reactive and prediction-based autoscaling, ScaleRAN dynamically reallocates GPU resources between latency-critical RAN processing and co-located AI workloads while preserving slot-level timing constraints. To enable realistic and reproducible evaluation, we also introduced LDPCSim, a GPU-accelerated workload generator built on NVIDIA Aerial for generating representative 5G NR LDPC decoding workloads. Together, these contributions demonstrate the feasibility of fine-grained GPU sharing for future AI-RAN platforms. 
The proposed design is well suited to O-RAN O-Cloud deployments, where elastic GPU sharing enables efficient AI-RAN co-location.

As future work, we plan to contribute to O-RAN WG6 discussions on AI-enabled O-Cloud platforms. We also intend to extend ScaleRAN beyond SM allocation by managing other resources, such as device memory. We will also investigate AI workload profiling techniques to characterize execution phases, kernel granularity, and SM footprint, enabling ScaleRAN to automatically synchronize profile transitions with kernel launch boundaries. This would improve support for AI models whose inference  exceeds the RAN slot duration. Finally, we will investigate a Kubernetes operator for managing the ScaleRAN runtime and workload placement. We are also considering open-sourcing LDPCSim and ScaleRAN.

\balance

\bibliographystyle{IEEEtran}
\bibliography{Biblio}

\end{document}